\documentclass[aps,prl,twocolumn,groupedaddress]{revtex4}

\usepackage{graphicx}

\begin{document}


\title{Dimensionality effects on non-equilibrium electronic transport 
in Cu nanobridges}


\author{D. Beckmann}
\email[e-mail address: ]{detlef.beckmann@int.fzk.de}
\author{H. B. Weber}
\affiliation{Forschungszentrum Karlsruhe, Institut f\"ur Nanotechnologie, P.O. Box 3640, D-76021 Karlsruhe, Germany}
\author{H. v. L\"ohneysen}
\affiliation{Forschungszentrum Karlsruhe, Institut f\"ur Festk\"orperphysik, P.O. Box 3640, D-76021 Karlsruhe, Germany,
and Physikalisches Institut, Universit\"at Karlsruhe, D-76128 Karlsruhe, Germany}


\date{\today}

\begin{abstract}
We report on non-equilibrium electronic transport through normal-metal (Cu)
nanobridges coupled to large reservoirs at low temperatures. 
We observe a logarithmic temperature dependence of the zero-bias conductance, 
as well as a universal scaling behavior of the differential conductance.
Our results are explained by electron-electron interactions in diffusive
metals in the zero-dimensional limit.
\end{abstract}


\maketitle

In clean bulk metals, the electron-electron interaction, although strong, is 
screened sufficiently to make an independent electron picture valid for
a broad range of materials. In disordered metals, however, screening is
limited by the diffusive motion of electrons, 
leading to a well-known correction to the linear conductance and to the
one-particle density of states (DOS) near the Fermi level \cite{aa}.
The corrections depend on dimensionality, and for the DOS are 
proportional to  $\sqrt\epsilon$, $\ln^2\epsilon$ and $\sqrt\epsilon^{-1}$ in 
three, two and one dimensions, respectively, where $\epsilon$ is the 
energy difference to the Fermi level, and logarithmic corrections to
leading power laws have been omitted. The DOS correction is 
spectroscopically resolved
as a zero-bias anomaly (ZBA) in tunneling experiments. For metallic point 
contacts, the occurrence of ZBAs has been reported in the literature for a 
long time \cite{yanson1986,ralph1994,anaya2003,yu2003},
and attributed to various physical mechanisms like Kondo scattering by
magnetic impurities, two-channel Kondo scattering
by two-level systems, or Coulomb interaction.

In our previous work \cite{weber2001}, we have presented transport measurements
on thin, short metallic films connected to large reservoirs.
In these films, all inelastic scattering lengths were much larger than the sample size,
such that a well-defined non-equilibrium distribution function with a double-step 
\cite{pothier1997} in the electronic system can be assumed. 
The elastic mean free path was of the order of the film thickness, while the
lateral dimensions of the films were much larger, i.e.
the diffusive random-walk of the electrons is essentially two-dimensional (2D).
We observed a logarithmic temperature dependence of the linear conductance,
and a ZBA in the differential conductance which obeyed a parameter-free 
scaling relation with a logarithmic
energy dependence. 
From a careful analysis of the energy and magnetic-field dependence and the 
reproducible amplitude of the ZBA, independent of sample preparation, we ruled out 
an explanation based on Kondo or two-channel Kondo physics for the 
anomaly observed in our samples. 
Our results were explained theoretically based on the
Aharonov-Altshuler corrections in two dimensions extended to the 
non-equilibrium situation in a finite-size sample. 
Recently, further theoretical work on Coulomb interaction in
non-equilibrium transport of metallic nanobridges 
\cite{golubev2001} has been reported.
A logarithmic correction to the conductance,
and a scaling form indistinguishable from our results, is predicted.
These results are found to be independent of microscopic details, 
as long as the sample is fully phase coherent. Especially, the predictions are
not specific to a two-dimensional thin film sample.
We have therefore extended our investigation to clarify 
the dependence of the logarithmic ZBA on dimensionality by changing the 
sample geometry.

\begin{figure}
\includegraphics[width=\columnwidth]{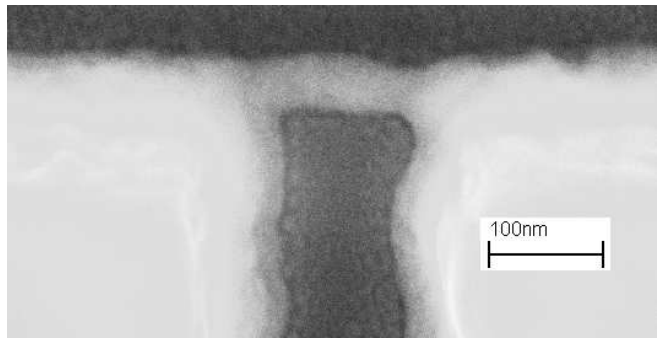}%
\caption{\label{sem}SEM image of sample Cu2. A thin bridge connects the two
massive reservoirs. Thus, a well-defined non-equilibrium situation can
be created inside the bridge by applying a bias voltage to the
reservoirs.}
\end{figure}


\begin{table}
\caption{\label{tab_exp}Characteristic parameters of our samples.
Geometry $\mathrm{width}~W\times\mathrm{height}~H\times\mathrm{length}~L$, 
resistance $R$, resistivity $\rho$, elastic mean free path $l$, 
diffusion constant $D$, and Thouless energy $E_\mathrm{Th}$ divided by the
Boltzmann constant $k_\mathrm{B}$.
}
\begin{ruledtabular}
\begin{tabular}{lcccccc}
sample & $W\times H\times L$ & $R$        & $\rho$                    & $l$     
       & $D$                         & $E_\mathrm{Th}/k_\mathrm{B}$ \\ 
       & $(\mathrm{nm}^3)$   & $(\Omega)$ & $(\mathrm{\mu\Omega cm})$ & $(\mathrm{nm})$ & $(\mathrm{cm}^2/\mathrm s)$ & $(\mathrm{K})$  \\ \hline
Cu1    & $25\times25\times160$ & 31.7  & 12.4 & 5.3  & 20.9  & 3.9   \\
Cu2    & $35\times25\times140$ & 16.7  & 10.4 & 6.3  & 24.8  & 6.1   \\
\end{tabular}
\end{ruledtabular}
\end{table}

The samples presented in this study were prepared by e-beam lithography
and shadow-evaporation techniques.
A silicon wafer with 600~nm of thermal oxide was coated with two
layers of e-beam resist, with the lower layer being PMMA/MAA and the upper
layer PMMA 950K. Then, a mask with large areas for the reservoirs and a 
thin slit for the bridge was defined by e-beam lithography. 250~nm of Cu
was evaporated under an angle of $13^\circ$ through the mask, forming the reservoirs,
contact leads and a shadow of the bridge unconnected to the rest
of the structure. Then, the bridge was evaporated under an
angle of $-10^\circ$ through the mask, closing the gap between the reservoirs.
Here, the thick Cu layer on top of the resist resulting from the first
evaporation somewhat narrows the slit left 
for the bridge, allowing for bridges down to approximately 20~nm width.
After lift-off the samples were quickly mounted into
a dilution refrigerator. Total exposure time to air was kept to less than 
30 minutes in order to minimize oxide formation.
The zero-bias resistance was measured using an AC resistance bridge. 
Care was taken
to keep the excitation amplitude below the thermal energy scale.
For the differential resistance 
measurements, an additional DC current was applied. Measurements were
 carried out at temperatures between
0.1 and 3~K and with magnetic fields of up to 12~T applied perpendicular 
to the substrate plane.
We present data of two samples, Cu1 and Cu2, with 
parameters given in Table \ref{tab_exp}.
The mean free path $l$ was obtained from $\rho l$, and the diffusion constant $D$ 
from Einstein's relation $\rho^{-1} = Ne^2D$, where $\rho$ is the resistivity
of the bridge, and $N$ the density of states at the Fermi level. 
The material parameters were taken from \cite{Ashcroft}.
From $D$, the Thouless energy $E_\mathrm{Th}=hD/L^2$ was calculated. 

\begin{figure}
\includegraphics[width=\columnwidth]{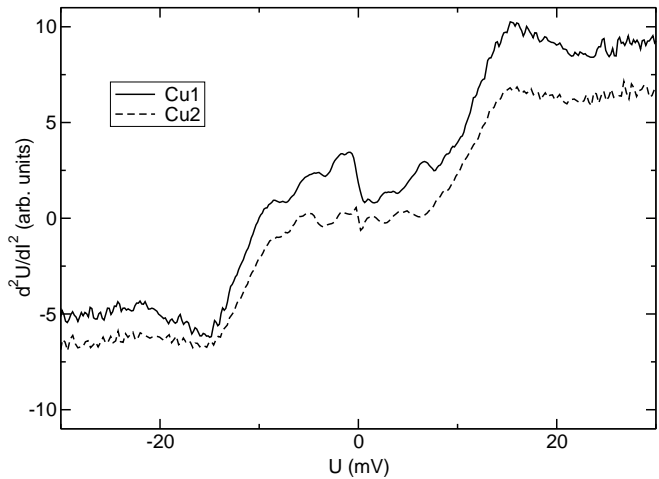}%
\caption{\label{fig_pcs}Point contact spectra of our samples. 
The data are scaled to a similar amplitude and offset for clarity.}
\end{figure}

To check whether the measurements indeed establish a well-defined 
non-equilibrium with negligible inelastic scattering, 
differential resistance measurements were carried
out to bias voltages up to 30~mV in order to observe
point contact spectra. These high-bias data
are plotted in Fig. \ref{fig_pcs} as $d^2U/dI^2$ vs.
$U$. The data clearly exhibit peaks at $\approx 15~\mathrm{mV}$, 
followed by a broad shoulder. These peaks can be related to backscattering
by optical phonons in the reservoirs, and are consistent with 
published results for copper
point contacts in the diffusive regime \cite{yanson1986}. 
The observation of well-resolved point contact spectra at bias voltages
an order of magnitude beyond the regime of the ZBA clearly shows that broadening 
of the energy distribution due to self-heating is small. 
Independently, we have estimated the effective temperature due to self-heating
in a model similar to that presented in Ref. \cite{henny1999}, and found
a negligible temperature rise compared to the bias voltage.

\begin{figure}
\includegraphics[width=\columnwidth]{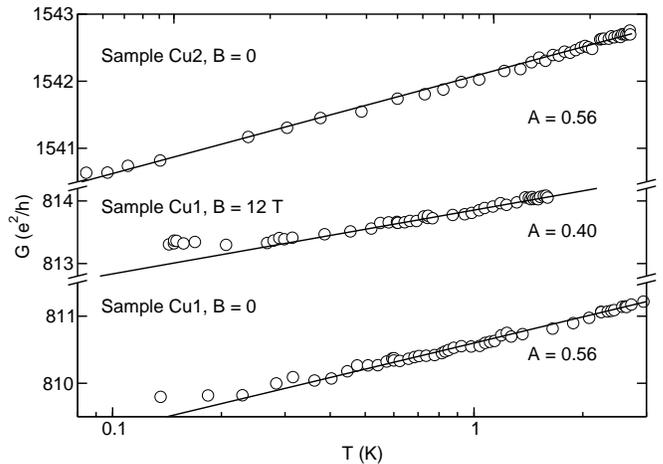}%
\caption{\label{cu_temp}Temperature dependence of the conductance of samples 
Cu1 at $B=0$ and $B=12~\mathrm{T}$ and Cu2 at $B=0$. 
The solid lines are fits to (\ref{equ_tdep}), with amplitudes
$A$ as shown.}
\end{figure}

In Fig. \ref{cu_temp}, the temperature dependence of the zero-bias conductance
of the two samples is shown.
Both display a logarithmic temperature dependence in a
temperature range from 0.2~K to 1.2~K. The data can be fitted with
\begin{equation}
G(0,T) = G(0,T_0) + A \ln(T/T_0), \\
\label{equ_tdep}
\end{equation}
where $G$ is the conductance, $T$ the temperature and $T_0$ an arbitrary reference
temperature. The amplitude $A$ in zero applied
magnetic field is about 0.56~$e^2/h$ for both samples. For sample
Cu1, measurements were performed in a magnetic field $B=12~{\rm T}$, where 
the amplitude is reduced to 0.40~$e^2/h$. 

\begin{figure}
\includegraphics[width=\columnwidth]{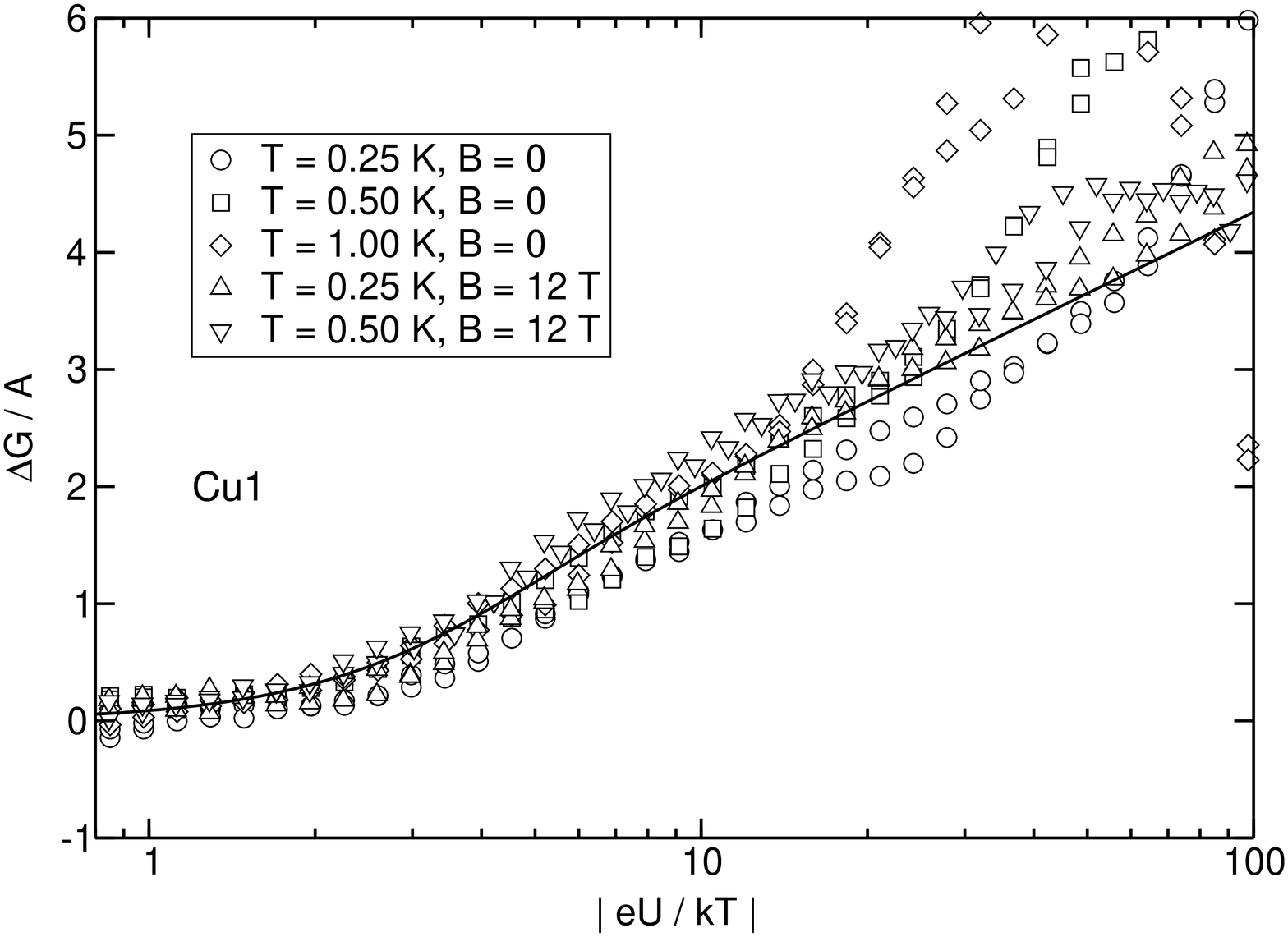}%
\caption{\label{cu_skal_a}Scaling plot of the differential conductance of sample Cu1 for
different temperatures and magnetic fields. 
The solid line is the scaling function $\Phi(eU/k_\mathrm BT)$.}
\end{figure}

\begin{figure}
\includegraphics[width=\columnwidth]{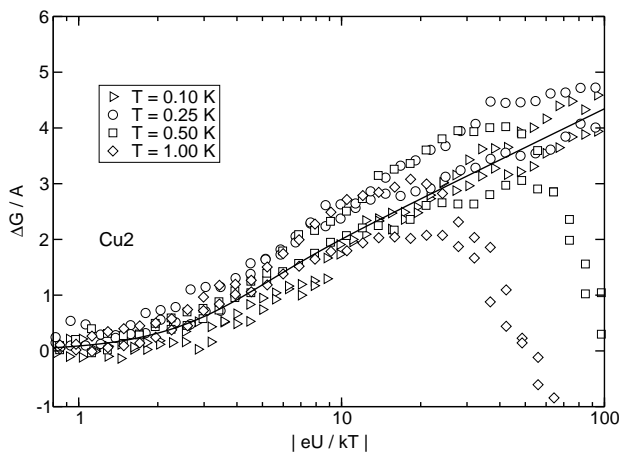}%
\caption{\label{cu_skal_b}Scaling plot of the differential conductance of 
sample Cu2 for different temperatures and magnetic field $B=0$. 
The solid line is the scaling function $\Phi(eU/k_\mathrm BT)$.}
\end{figure}

The scaling form
\begin{equation}
\frac{G(U,T)-G(0,T)}{A} = \Phi\left( \frac{eU}{k_{\rm B}T} \right) \\
\label{equ_scaling}
\end{equation}
is employed to describe our data \cite{weber2001}. Here, $U$ is the applied voltage, $k_{\rm B}$ the
Boltzmann constant, and $\Phi$ is a universal function. This formula 
suggests rescaling
and plotting the data as $(G(U)-G(U=0))/A$ vs. $eU/k_{\rm B}T$. 
Such scaling plots of the data for samples Cu1 and Cu2 are shown in Figs. 
\ref{cu_skal_a} 
and \ref{cu_skal_b}. As can be seen, the data of both samples collapse onto
a common function $\Phi$ in the low bias regime, whereas for larger bias, 
deviations with a reproducible dependence on $eU$ instead of $eU/k_{\rm B}T$ are 
observed. The onset of these deviations
is consistent with our estimate of the Thouless energy. These deviations 
are presumably voltage-dependent universal conductance fluctuations (VUCF), 
and can also 
be seen as small wiggles in the point contact spectra in Fig. \ref{fig_pcs}, where 
the ZBA itself is resolved as a steep negative slope around zero bias.

The universal scaling function $\Phi$ 
has been calculated by means of a tunneling Hamiltonian, which couples
the interacting diffusive metal to ideal leads \cite{weber2001}. 
As the bridge is shorter than all inelastic scattering length scales,
the bias voltage is expected to create a non-equilibrium double-step 
distribution 
function in the electronic system, leading to a spectroscopic resolution
of interaction corrections to the density of states.
It was shown that the $\ln^2\epsilon$ dependence
of two-dimensional films is effectively cut off to a simple
$\ln\epsilon$ dependence, as 
only the lowest lying diffusion mode is relevant for our finite-length samples
as long as
energies below the Thouless energy are considered. 
A similar treatment \cite{schwab2002} confirms this result, and predicts the
amplitude $A$ of the anomaly to be in the range $0.25-0.5~e^2/h$, depending
on the transparency of the bridge-to-lead interface.

An alternative treatment of the ZBA in a coherent conductor in series with
an environmental impedance has been given by Golubev et al. \cite{golubev2001} 
and Yeyati et al. \cite{yeyati2001}. In this case, the ZBA is interpreted 
as a residue 
of the well known Coulomb blockade in the limit of high transparency.
The theory does not depend on microscopic details or the extent of the
conductor, as long as it is phase-coherent. A logarithmic temperature 
dependence is predicted, along with a scaling function that is 
numerically indistinguishable 
from (\ref{equ_scaling}) \cite{golubev2001}. 
The amplitude $A$ is predicted to be 2/3~$e^2/h$ for a diffusive conductor, in which
case the resistance of the conductor itself serves as the environmental 
impedance.

For our present study, 
we deliberately varied the dimensionality of the diffusive motion
compared to our previous work: Width, height and mean free path
are approximately
equal, thus the nanobridges are now short wires (1D)
instead of films (2D).
Nonetheless, the logarithmic temperature dependence and 
scaling behavior calculated for the short films is observed,
with an amplitude $A\approx 0.4-0.5~e^2/h$ in  agreement 
with the range of the predicted values \cite{golubev2001,schwab2002}.
We find that the scaling behavior is valid in the 
bias voltage range defined by the Thouless energy.
We therefore conclude that the lateral dimensions of the sample
compared to the mean free path, and therefore the dimensionality of
the diffusive motion, are irrelevant,
as long as $W,H,L<\sqrt{hD/\max(eU,k_\mathrm{B}T)}$, and only the lowest
lying diffusion mode is realized. Thus, even though it employs
a film or wire geometry, our experiment effectively constitutes the
zero-dimensional limit of electron-electron interaction. This fact has been
pointed out by a theoretical treatment similar to our previous work
\cite{schwab2002}. 
The consistency with the results of Golubev et al. \cite{golubev2001}
demonstrates the close relationship between the environmental Coulomb interaction
and the electron-electron interaction in diffusive metals, which has recently 
been demonstrated in a tunneling experiment \cite{pierre2001}.


To conclude, we have presented transport measurements on short, 
phase-coherent metallic
wires in a well-defined electronic non-equilibrium situation. 
We find the same logarithmic temperature dependence and scaling
behavior as observed previously in short films. We therefore
identify our results with electron-electron interactions in
a diffusive metal in the zero-dimensional limit. Our experiment
demonstrates the close relation between electron-electron interaction
in diffusive metals and the environmental Coulomb interaction.

\begin{acknowledgments}
Useful discussions with P. Schwab, J. Kroha and A. Zaikin are gratefully
acknowledged, as well as the support of M. Br\"uckel in the sample fabrication.
\end{acknowledgments}

\bibliography{../lit}

\end{document}